\documentclass{camera}
\usepackage{graphicx}

\begin{document}

\title{A comparison of the fluences and photon\\
       peak fluxes for the \emph{Swift} and \emph{RHESSI}\\
       gamma-ray bursts}

\author{D. Huja$^{1}$, J. \v{R}\'{\i}pa$^{1}$ \and A. M\'esz\'aros$^{1}$}
\organization{
$^{1}$ Charles University, Faculty of Mathematics and Physics,\\
Astronomical Institute, Prague, Czech Republic}
\maketitle

\begin{abstract}
Fluences and photon peak fluxes of the gamma-ray bursts (GRBs)
detected by the \emph{Swift} and \emph{RHESSI} satellites are graphically compared.
\end{abstract}

The \emph{Swift}-BAT data set consists of the $T_{90}$ GRB durations, fluences at range 15-150\,keV,
peak fluxes (fluxes) at the same energy range and covers the period Nov. 2004 – Feb. 2009 with 429 events.
The \emph{RHESSI} data set consists of the $T_{90}$ durations, count fluences at range 25-1500\,keV,
fluxes at the same range and covers the period Feb. 2002 – Nov. 2007 with 425 events.
In these two databases 25 GRBs were detected by both satellites with known fluxes and fluences.
The relationship between the \emph{RHESSI} fluxes (counts/s) and the \emph{Swift} fluxes (ph\,cm$^{-2}$s$^{-1}$),
and the relationship between the \emph{RHESSI} fluences (counts) and the \emph{Swift} fluences (10$^{-7}$\,erg\,cm$^{-2}$)
are plotted in fig.~1. There is no simply relation to convert the \emph{RHESSI} count fluxes/fluences to the \emph{Swift} fluxes/fluences
in energy units, but there exists a very strong correlation between them.
The \emph{Swift} fluxes of the short and the long GRBs are on average approximately the same, but for the \emph{RHESSI} data
the fluxes of the short GRBs are on average higher than for the long ones.
In the case of fluences, for both samples, the long GRBs have on average higher values than the short ones.
Also the gradient of the best-fitted straight line in the fluence-$T_{90}$ dependency is very similar for both data sets.
\newline

We acknowledge support of the grants GAUK 46307, GA\v{C}R
205/08/H005, MSM0021620860, and OTKA K077795.

\begin{figure}
\begin{center}
\includegraphics[width=0.95\textwidth]{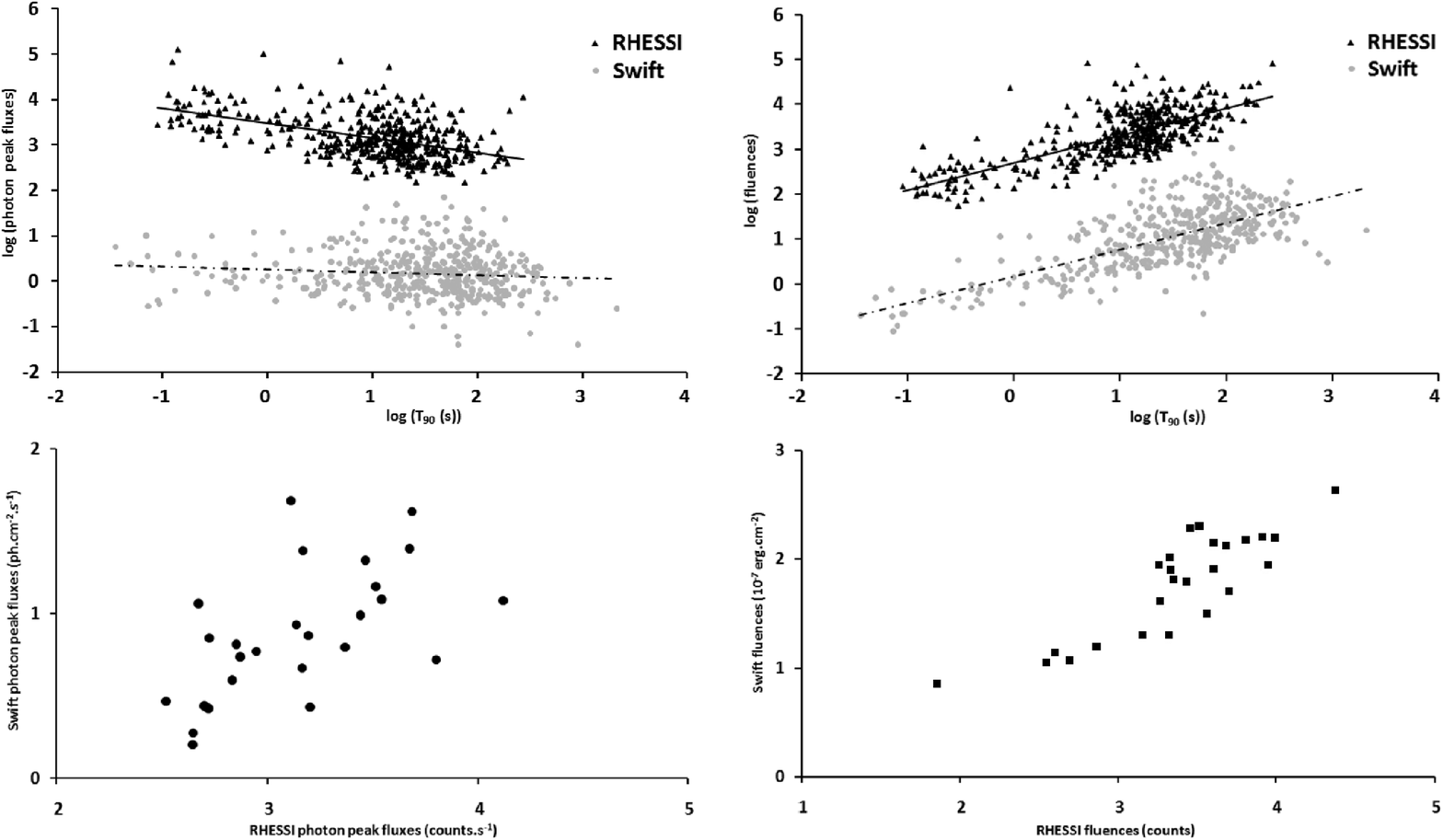}
\caption{
\emph{Top left:} Duration vs. fluxes: Fluxes measured by \emph{Swift} are on average the same for the short and long GRBs,
the gradient of the best-fitted straight line is -0.0656 \cite{ref01}, but \emph{RHESSI} fluxes are on average higher
for the short GRBs than for the long ones. The gradient of this relation is -0.3255 \cite{ref01}.
\emph{Top right:} Duration vs. fluences: Fluences measured by \emph{Swift} are on average higher for the long GRBs than for the short ones,
the gradient is 0.6077 \cite{ref01}. Fluences obtained by \emph{RHESSI} for the long GRBs are on average higher than for the short ones
with similar gradient: 0.5930 \cite{ref01}.
\emph{Bottom left:} The relationship between fluxes detected by \emph{RHESSI} and \emph{Swift} for the common 25 GRBs.
The correlation coefficient is 0.5840 \cite{ref02}, t-coeff. is 3.450
(in the null case of no correlation like Student's t-distribution with 25-2=23 dof),
probability of no correlation is not higher than 0.2\,\%.
\emph{Bottom right:} The relationship between the \emph{RHESSI} and \emph{Swift} fluences for the common 25 GRBs.
The correlation coeff. is 0.8511 \cite{ref02}, t-coeff. is 7.776, probab. of no correlation is not higher than 7x10$^{-6}$\,\%.}
\end{center}
\end{figure}

\end{document}